\documentclass[sigconf]{acmart}
\usepackage{multirow}
\usepackage{graphicx}
\usepackage{svg}
\usepackage{caption}
\usepackage{mathtools}
\usepackage{relsize}
\usepackage{balance}
\AtBeginDocument{%
  \providecommand\BibTeX{{%
    \normalfont B\kern-0.5em{\scshape i\kern-0.25em b}\kern-0.8em\TeX}}}

\setcopyright{acmlicensed}
\copyrightyear{2018}
\acmYear{2018}
\acmDOI{XXXXXXX.XXXXXXX}

\copyrightyear{2024}
\acmYear{2024}
\setcopyright{acmlicensed}
\acmConference[SIGIR '24] {Proceedings of the 47th International ACM SIGIR Conference on Research and Development in Information Retrieval}{July 14--18, 2024}{Washington, DC, USA.}
\acmBooktitle{Proceedings of the 47th International ACM SIGIR Conference on Research and Development in Information Retrieval (SIGIR '24), July 14--18, 2024, Washington, DC, USA}
\acmISBN{979-8-4007-0431-4/24/07}
\acmDOI{10.1145/3626772.3661356}

\begin{document}

\title{A Unified Search and Recommendation Framework Based on Multi-Scenario Learning for Ranking in E-commerce}


\author{Jinhan Liu}
\authornote{Corresponding author}
\affiliation{%
  \institution{JD.com}
  \city{Beijing}
  \country{China}
}
\email{liujinhan1@jd.com}

\author{Qiyu Chen}
\affiliation{%
  \institution{JD.com}
  \city{Beijing}
  \country{China}
}
\email{chenqiyu@jd.com}

\author{Junjie Xu}
\affiliation{%
  \institution{JD.com}
  \city{Beijing}
  \country{China}
}
\email{xujunjie33@jd.com}

\author{Junjie Li}
\affiliation{%
  \institution{JD.com}
  \city{Beijing}
  \country{China}
}
\email{lijunjie72@jd.com}

\author{Baoli Li}
\affiliation{%
  \institution{JD.com}
  \city{Beijing}
  \country{China}
}
\email{libaoli5@jd.com}

\author{Sulong Xu}
\affiliation{%
  \institution{JD.com}
  \city{Beijing}
  \country{China}
}
\email{xusulong@jd.com}

\renewcommand{\shortauthors}{Jinhan Liu et al.}


\begin{abstract}
Search and recommendation (\textbf{S\&R}) are the two most important scenarios in e-commerce. The majority of users typically interact with products in S\&R scenarios, indicating the need and potential for joint modeling. Traditional multi-scenario models use shared parameters to learn the similarity of multiple tasks, and task-specific parameters to learn the divergence of individual tasks. This coarse-grained modeling approach does not effectively capture the differences between S\&R scenarios. Furthermore, this approach does not sufficiently exploit the information across the global label space. These issues can result in the suboptimal performance of multi-scenario models in handling both S\&R scenarios. 
To address these issues, we propose an effective and universal framework for \textbf{U}nified \textbf{S}earch and \textbf{R}ecommendation (USR), designed with \textit{S\&R Views User Interest Extractor Layer} (IE) and \textit{S\&R Views Feature Generator Layer} (FG) to separately generate user interests and scenario-agnostic feature representations for S\&R. Next, we introduce a \textit{Global Label Space Multi-Task Layer} (GLMT) that uses global labels as supervised signals of auxiliary tasks and jointly models the main task and auxiliary tasks using conditional probability. Extensive experimental evaluations on real-world industrial datasets show that USR can be applied to various multi-scenario models and significantly improve their performance. Online A/B testing also indicates substantial performance gains across multiple metrics. Currently, USR has been successfully deployed in the 7Fresh App.

\end{abstract}
\vspace{-0.2cm}


\maketitle
\vspace{-0.2cm}
\section{Introduction}
Search and recommendation functions have become essential services for online applications, with many researchers working to advance the outcomes of these services \cite{din,dien,dmr,TWIN,widedeep,deepfm,xDeepFM}. In addition, many studies \cite{sr1,sr2,sr3,sr4,sr5} have validated that the S\&R scenarios complement each other in terms of their functions and information. Integrating S\&R into a joint modeling approach is a critical and valuable problem to solve. In fact, some state-of-the-art work has garnered impressive results in Multi-Scenario Learning (MSL). Based on the different architectures, the existing Multi-Scenario Models (MSM) can be divided into two categories: Tower-based models \cite{hmoe,mmoe,ple,AESM2,hinet} and Dynamic Weight (DW) models \cite{apg,adasparse,pepnet}. These studies typically focus on modeling scenarios that are similar, such as different recommendation locations. However, when applied to S\&R scenarios, the inherent limitations of traditional multi-scenario modeling methods may lead to less than optimal outcomes. Traditional MSM are unable to extract scenario-specific representations from the combined behavior sequences collected from both S\&R scenarios. For domain-agnostic features, MSM do not differentiate their importance across the respective scenarios, treating them indiscriminately. Furthermore, traditional MSM do not utilize cross-domain label information, which could serve as data augmentation for positive samples in each scenario. Therefore, these potential gains have been neglected.
\begin{figure*}[!ht]
\centering
\includegraphics[width=\linewidth]{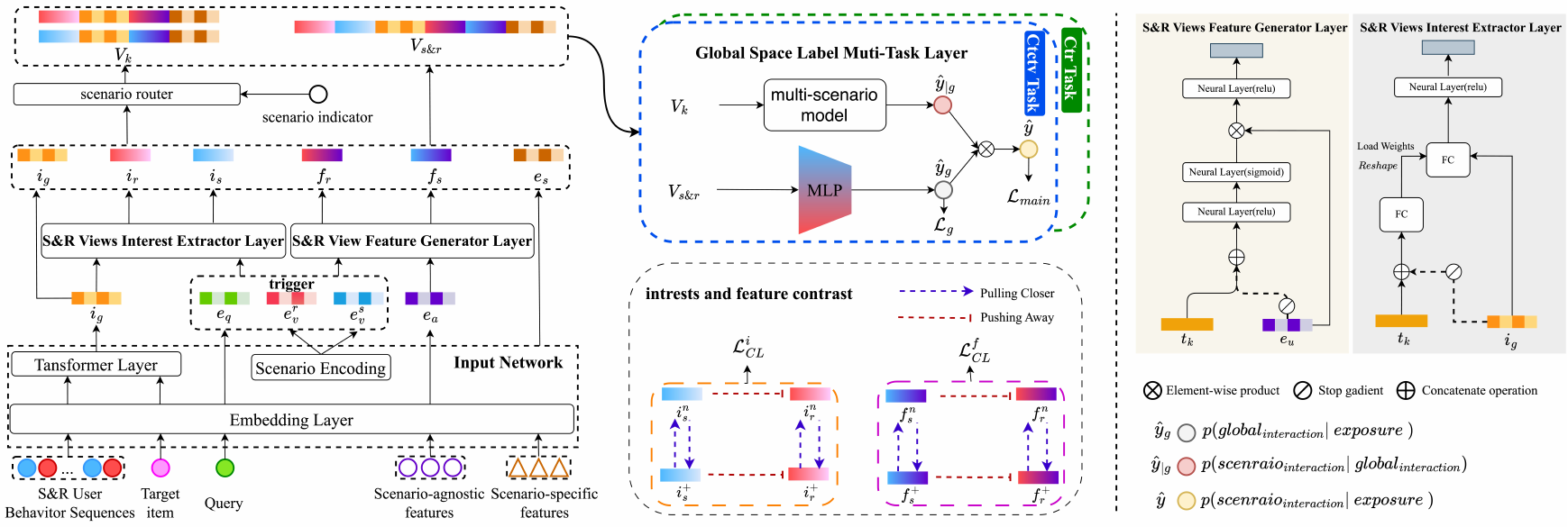}
\caption{Overall structure of USR.}
\label{fig:model}
\vspace{-0.3cm}
\end{figure*}
To this end, in this paper, we propose a Unified Search and Recommendation (USR) framework based on MSL. In the bottom of USR, we design User Interest Extractor Layer (IE) and Feature Generator Layer (FG) to obtain fine-grained representations for user interests and scenario-agnostic features in S\&R views. Specifically, we define a scenario trigger that introduces prior knowledge of the scenario. After that, we extract user interests and scale domain-agnostic features across S\&R views using dynamic weight technique and gating mechanism with the S\&R trigger. Moreover, we apply contrastive learning to these representations to further capture the differences between S\&R scenarios. In the Global Label Space Multi-Task Layer (GLMT), we propose a novel method that jointly models scenario-specific CTR/CTCVR and global CTR/CTCVR, based on conditional probability. The purpose is to let the model perceive global label information during learning. To summarize, the contributions of this paper are as follows:

\begin{itemize}
    \item To the best of our knowledge, this is the first work that improves upon multi-scenario learning for the joint modeling of search and recommendation.
    \item We propose a USR framework based on MSL to enhance search and recommendation. Through the IE and FG, we obtain fine-grained representations of user interests and scenario-agnostic features within S\&R views. Additionally, we design the GLMT which employs global labels as supervised signals for auxiliary tasks and jointly models the main and auxiliary tasks using conditional probability.   
    \item Both offline and online experiments demonstrate the superiority of our proposed USR. Currently, USR has been successfully deployed in the 7Fresh\footnote{7Fresh (https://www.7fresh.com) is an online-to-offline (O2O) service platform provided by JD.com Group.} App.
\end{itemize}
\vspace{-0.2cm}
\section{PROBLEM FORMULATION}
Considering a training dataset $\mathcal{D}=\left\{\left(x_i, y_i^j\right)\right\}_{i=1}^{|\mathcal{D}|}$, where $x_i$ represents the feature set and $y_i^j \in\{0,1\}$ represents the class label indicating a click for click-through rate (CTR) prediction or a purchase for click-through \& conversion rate (CTCVR) prediction of the $i_{t h}$ sample. Note that the original input features include: S\&R user behavior sequence $B_{S\&R}$, query $q$, target item $i$, scenario indicator $k$ and other features $x$. In addition, we divide the other features into two parts based on whether they are scenario-aware. We need to train a unified model, to predict multiple tasks in S\&R scenarios, which can be formulated as:
\begin{equation}
\hat{y}^{j}= \begin{cases}f_{\Theta}(B_{S\&R},i,k,x_s,x_a) & { if \: k=r } \\ f_{\Theta}(B_{S\&R},i,k,q,x_s,x_a) & { if \: k=s }\end{cases},
\end{equation}
$f_{\Theta}(\cdot)$ denotes the underlying unified model with parameters $\Theta$, $x_s$ is the scenario-specific feature set, $x_a$ is the scenario-agnostic feature set and $\hat{y}^{j}$ is the model’s prediction of the task-$j$.
\section{METHOD}
The network architecture of USR is illustrated in Fig. \ref{fig:model}. In this section, we present the proposed USR approach in detail.
\vspace{-0.2cm}
\subsection{The Input Network}
The embedding layer is shared across S\&R scenarios and learns the representation of the inputs. After transforming all features into embeddings, we denote the representations of S\&R user behavior sequence, target item, query, scenario-specific, and scenario-agnostic features as $\boldsymbol{b}_{S\&R}$, $\boldsymbol{e}_t$, $\boldsymbol{e}_q$, $\boldsymbol{e}_s$ and $\boldsymbol{e}_a$, respectively. Next, in the sequence encoding, we follow BST \cite{bst} to utilize $\boldsymbol{b}_{S\&R}$ and $\boldsymbol{e}_t$ to extract the global user interests, denoted as $\boldsymbol{i}_g$. In the scenario encoding, inspired by \cite{mran}, we embed the S\&R scenario in the same space. Note that the scenario representation we extract is global rather than local. For each instance, there will be both S\&R views representation simultaneously. Let $\boldsymbol{e}_{v}^{s}$ and $\boldsymbol{e}_{v}^{r}$ denote S\&R views representations, respectively. Each of the embeddings previously mentioned has a dimension of $\boldsymbol{d}$. Additionally, we define a scenario trigger that introduces prior knowledge of scenario and influence the weight of feature information. The trigger $\boldsymbol{t}$ can be formulated as follows:
\begin{equation}
t_k= \begin{cases}e_v^r, & {  if \: k=r  } \\ e_v^s + e_q, &{ if \: k=s }\end{cases}.
\vspace{-0.2cm}
\end{equation}
\vspace{-0.3cm}
\subsection{S\&R Views User Interest Extractor Layer}
The global user interest representations obtained merely through sequence encoding do not effectively distinguish between user interest representations in S\&R scenarios. To address this issue, inspired by the Dynamic Weight (DW) technique \cite{apg}, we propose a layer to extract user interests for S\&R views based on the re-parameterization method to adaptively generate parameters depending on the given condition. Specifically, we concatenate the global user interests $\boldsymbol{i}_g$ and the scenario trigger $\boldsymbol{t}_k$ as the condition, with $\boldsymbol{i}_g$ as the input for the DW network. The network weights are adaptively generated under this condition, using Fully Connected (FC) layers and a \textit{Reshape} function. For each sample, given the scenario trigger $\boldsymbol{t}_k$, the user interest representations for S\&R views can be represented as:
\begin{equation}
i_s=f_{M L P}(\textit{DW}(t_s \oplus(\oslash i_g), i_g)),
\end{equation}
\begin{equation}
i_r=f_{M L P}(\textit{DW}(t_r \oplus(\oslash i_g), i_g)),
\end{equation}
where $\oplus$ indicates concatenation, $\oslash$ means stop gradient, $\boldsymbol{i}_s,\boldsymbol{i}_r \in \mathbb{R}^d$ denote the user interest representations for S\&R views, respectively. To better capture the differences in user interests between S\&R scenarios, we draw inspiration from classic \textit{contrastive learning} \cite{cl1,cl2,cl3}, and design a novel contrastive loss, which is formulated as follows:
\begin{equation}
\mathsmaller{\mathcal{L}_{CL}^i=-\underset{n}{\mathbb{E}}[\log \frac{\exp ( i_s^n \cdot i_s^{+} / \tau)+\exp ( i_r^n \cdot i_r^{+} / \tau)}{\exp ( i_s^n \cdot i_s^{+} / \tau)+\exp ( i_r^n \cdot i_r^{+} / \tau) + \sum_{k=1}^N \exp ( i_s^n \cdot i_r^k) / \tau)}]},
\label{eq:cl_i}
\end{equation}where $i_r^{+}$ and $i_s^{+}$ denote the positive samples, which are randomly sampled from $\boldsymbol{i}_r$ and $\boldsymbol{i}_s$ within a batch, $i_r^k$ denotes the negative sample, $N$ is the batch size, and $\tau$ is a temperature coefficient.
\vspace{-0.2cm}
\subsection{S\&R Views Feature Generator Layer}
Considering the heterogeneity of S\&R, it is apparent that scenario-agnostic features have different discrimination in S\&R. For example, product click-related features generally play a more significant role in the recommendation scenario than in the search scenario. Thus, FG is designed to scale each feature according to the S\&R trigger. Specifically, for each sample, we concatenate the scenario trigger $\boldsymbol{t}_k$ and the scenario-agnostic features $\boldsymbol{e}_a$ as the prior information to scale $\boldsymbol{e}_a$. Then, the dimension of the hidden vector is projected to $d$ dimensions through a feed-forward layer. Formally, we have the following equation:
\begin{equation}
f_s= f_{M L P}(e_a \cdot sigmoid(f_{M L P}(t_s \oplus(\oslash e_a))))
\end{equation}
\begin{equation}
f_r= f_{M L P}(e_a \cdot sigmoid(f_{M L P}(t_r \oplus(\oslash e_a))))
\end{equation}where $\boldsymbol{f}_s,\boldsymbol{f}_r \in \mathbb{R}^d$ denote the feature representations for the S\&R views, respectively, $sigmoid$ is the sigmoid activation function.
Just as we learned the representations for $\boldsymbol{i}_r$ and $\boldsymbol{i}_s$, we also apply the contrastive loss to $\boldsymbol{f}_r$ and $\boldsymbol{f}_s$, which is formulated as follows:
\begin{equation}
\mathsmaller{\mathcal{L}_{CL}^f=-\underset{n}{\mathbb{E}}[\log \frac{\exp ( f_s^n \cdot f_s^+ / \tau)+\exp ( f_r^n \cdot f_r^+ / \tau)}{\exp ( f_s^n \cdot f_s^+ / \tau)+\exp ( f_r^n \cdot f_r^+ / \tau) + \sum_{k=1}^N \exp ( f_s^n \cdot f_r^k) / \tau)}]},
\label{eq:cl_f}
\end{equation}
where $f_r^+$ and $f_s^+$ denote the positive samples, which are randomly sampled from $\boldsymbol{f}_r$ and $\boldsymbol{f}_s$ within a batch, $f_r^k$ denotes the negative sample, $N$ is the batch size, and $\tau$ is a temperature coefficient.

\vspace{-0.2cm}
\subsection{Global Label Space Multi-Task Layer}
In the multi-task layer, to fully utilize label information in S\&R, we propose a novel method that jointly models the main task and auxiliary tasks using conditional probability. The main task and auxiliary tasks are scenario-CTR/CTCVR prediction and global-CTR/CTCVR prediction, respectively. We construct a global label denoted as $y_g$, for a given user $u$ and product $p$, which is formulated as follows:
\vspace{-0.1cm}
\begin{equation}
y_{g \mid u, p}= \begin{cases}1, & { if \: y_{s \mid u, p} =1 \: or \: y_{r \mid u, p} =1 } \\ 0, & \text { else }\end{cases},
\end{equation}
where $y_{s}$, $ y_{r}$ are the labels of S\&R scenario, respectively.
In Fig. \ref{fig:model}, we have defined the probability of global action (i.e., click and order) as $\hat{\boldsymbol{y}}_{g}:p(global_{action} \mid exposure)$ , the probability of action in current scenario given global action as $\hat{\boldsymbol{y}}_{\mid g}:p(scenario_{action} \mid global_{action})$, and the probability of action in current scenario as $\hat{\boldsymbol{y}}:p(scenario_{action} \mid exposure)$. It is obvious that $\hat{\boldsymbol{y}} = \hat{\boldsymbol{y}}_{\mid g} * \hat{\boldsymbol{y}}_{g}$. The inputs are differentiated as $\boldsymbol{V}_k$ for the main task and $\boldsymbol{V}_{S\&R}$ for the auxiliary task, respectively. Both $\boldsymbol{V}_k$ and $\boldsymbol{V}_{S\&R}$ include representations of global user interests and scenario-specific features. The difference is that $\boldsymbol{V}_k$ only contains representations of user interest and domain-agnostic features from the current scenario view. However, $\boldsymbol{V}_{S\&R}$ includes both S\&R views representations, aiming to better predict the global label:
\begin{equation}
V_k= \begin{cases} {[i_g \oplus i_s \oplus f_s \oplus e_s]}, & { if \: k=S } \\ {[i_g \oplus i_r \oplus f_r \oplus e_s]}, & { if \: k=R }\end{cases}, 
\vspace{-0.2cm}
\end{equation}

\begin{equation}
V_{S\&R}= [i_g \oplus i_s \oplus f_s \oplus i_r \oplus f_r \oplus e_s] ,
\end{equation}where $k$ is the scenario indicator. We employ Multi-Layer Perception (MLP) as the main task network and Multi-Scenario Model (MSM) as the auxiliary task network. Then $\hat{\boldsymbol{y}}_{\mid g}$ and $\hat{\boldsymbol{y}}_{g}$ can be formulated as follows:
\begin{equation}
\hat{y}_{\mid g} = f_{M L P}(V_k), \qquad \hat{y}_{g} = f_{M S M}(V_{S\&R}).
\end{equation}
\vspace{-0.6cm}
\subsection{Model Optimization}
The objective functions applied in main task and auxiliary task is the cross entropy loss function, defined as:
\vspace{-0.1cm}
\begin{equation}
\mathcal{L}_{main}=\sum_{\mathbf{x} \in \mathcal{S} \cup \mathcal{R}} \sum_{j=1}^k \ell(y_{\mathbf{x}}^j,\hat{y}_{\mathbf{x}}^j), 
\end{equation}
\begin{equation}
\mathcal{L}_{aux}=\sum_{\mathbf{x} \in \mathcal{S} \cup \mathcal{R}} \sum_{j=1}^k \ell(y_{g_\mathbf{x}}^j,\hat{y}_{g_\mathbf{x}}^j),
\end{equation}where $\mathcal{S}$ is the set of search samples, $\mathcal{R}$ is the set of recommendation samples, $\ell(\cdot)$ is the cross-entropy loss function. For sample $\mathbf{x}$ in task-$j$, $\hat{y}_{\mathbf{x}}^j$ and $\hat{y}_{g_\mathbf{x}}^j$ are the predictions for the main and auxiliary tasks, $y_{\mathbf{x}}^j$ and $y_{g_\mathbf{x}}^j$ are the label and global label, respectively. Then the final objective function of our proposed USR is:
\begin{equation}
\mathcal{L}_{final}=\lambda_1 * \mathcal{L}_{main} + \lambda_2 * \mathcal{L}_{aux} + \lambda_3 * \mathcal{L}_{i} + \lambda_4 * \mathcal{L}_{f},
\end{equation}where $\lambda_i$ is a hyper-parameter to control the weight of loss and $\sum_{i=1}^4 \lambda_i=1$, $\mathcal{L}_{i}$ and $\mathcal{L}_{f}$ are the contrastive loss by from Equation \ref{eq:cl_i} and \ref{eq:cl_f}.
\begin{table*}[h]
\caption{The AUC (\%) results of CTR and CTCVR prediction within S\&R scenarios. Note Base refers to the original results of the corresponding methods and Base+USR refers to the results with the help of USR. $\Delta$ refers the improvement.}
\label{tb:res}
\resizebox{\textwidth}{!}{
\begin{tabular}{c|l|c c|c c|c c|c c| cc| cc}
\toprule
 \multirow{2}{*}{\textbf{Scenario}} & \textbf{Method} & \multicolumn{2}{|c|}{ \textbf{ShareBottom} } & \multicolumn{2}{|c|}{ \textbf{MMoE} } & \multicolumn{2}{c|}{ \textbf{PLE} } & \multicolumn{2}{c|}{ \textbf{STAR} } & \multicolumn{2}{c|}{ \textbf{M2M} } & \multicolumn{2}{c}{ \textbf{APG} }\\
\cmidrule(){2-14}

\multirow{6}{*}{S}
& Metric &  $\mathrm{AUC_{CTCVR}}$ & $\mathrm{AUC_{CTR}}$ &  $\mathrm{AUC_{CTCVR}}$  &  $\mathrm{AUC_{CTR}}$ &  $\mathrm{AUC_{CTCVR}}$ &  $\mathrm{AUC_{CTR}}$ &
 $\mathrm{AUC_{CTCVR}}$ &  $\mathrm{AUC_{CTR}}$ &  $\mathrm{AUC_{CTCVR}}$ &  $\mathrm{AUC_{CTR}}$ &  $\mathrm{AUC_{CTCVR}}$ &  $\mathrm{AUC_{CTR}}$ \\
\midrule
& Base & 0.8298 & 0.8227 & 0.8327 & 0.8250 & 0.8331 & 0.8257 & 0.8330 & 0.8259 & 0.8316 & 0.8231 & 0.8337 & 0.8261\\
& Base+USR & \textbf{0.8347} & \textbf{0.8268} & \textbf{0.8352} & \textbf{0.8270} & \textbf{0.8359} & \textbf{0.8271} & \textbf{0.8358} & \textbf{0.8274} & \textbf{0.8343} & \textbf{0.8266} & \textbf{0.8353} & \textbf{0.8269}\\
& $\Delta$ & +0.49 & +0.41 & +0.25 & +0.20 & +0.28 & +0.14 & +0.28 & +0.15 & +0.27 & +0.35 & +0.16 & +0.08 \\
\midrule
\multirow{3}{*}{R} 
& Base & 0.6894 & 0.6529 & 0.6909 & 0.6541 & 0.6917 & 0.6568 & 0.6921 & 0.6524 & 0.6904 & 0.6527 & 0.6919 & 0.6561\\
& Base+USR & \textbf{0.6951} & \textbf{0.6569} & \textbf{0.6960} & \textbf{0.6578} & \textbf{0.6979} & \textbf{0.6577} & \textbf{0.6977} & \textbf{0.6581} & \textbf{0.6963} & \textbf{0.6571} & \textbf{0.6981} & \textbf{0.6584}\\
& $\Delta$ & +0.57 & +0.40 & +0.51 & +0.37 & +0.62 & +0.09 & +0.56 & +0.17 & +0.59 & +0.44 & +0.62 & +0.23 \\

\bottomrule
\end{tabular}}
\vspace{-0.3cm}
\end{table*}
\vspace{-0.3cm}
\section{EXPERIMENT}
\subsection{Experimental Setup}
\subsubsection{\textbf{Dataset.}} 
We collected an industrial dataset from the traffic logs of the 7Fresh S\&R system, spanning from Oct. 1 to Dec. 1, 2023. The training dataset includes over 5,304,568 search sessions and 540,471 recommendation sessions. For each session, items that were purchased or clicked were labeled as positive examples for the CTCVR or CTR tasks, and all items that were shown but not purchased or clicked were labeled as negative examples. 
\vspace{-0.1cm}
\subsubsection{\textbf{Baselines.}} 
To show the effectiveness of the proposed USR framework, we apply it to the following Multi-Scenario Learning (MSL) methods: 1) SharedBottom \cite{SB}, which adopts the multi-task learning framework that shares the parameters of the bottom layer and designs scenario-specific parameters for each scenario. 2) MMoE \cite{mmoe}, which designs the shared multi-gate mixture experts structure to model task relationships and capture common representations from multiple scenarios. 3) PLE \cite{ple}, which uses a progressive layered extraction for multi-task learning. Similar to MMoE, we apply scenario-specific experts and towers for each scenario. 4) STAR \cite{star}, a novel model with star topology that can cater to multiple domains using a unified model. 5) M2M \cite{m2m}, which designs a meta unit to generate specific parameters to capture scenario-specific knowledge. 6) APG \cite{apg}, which generates domain-specific parameters using additional networks and uses low-rank decomposition to reduce computation cost.
\vspace{-0.2cm}
\subsubsection{\textbf{Evaluation metrics and hyper-parameters.}} 
To evaluate the performance of all methods, we adopt the commonly-used AUC (Area Under ROC) as the metric. For a fair comparison, all methods use the same feature embedding size $d$ of 32, and employ the AdamW \cite{adamw} optimizer with an initial learning rate of 1E-4 and a batch size of 1024. The hyper-parameters in USR are set as follows: $\tau=1$, $\lambda_1=0.9$, $\lambda_2=0.08$, $\lambda_3=0.01$, $\lambda_4=0.01$. The output sizes of hidden layers in the auxiliary task are set to \{128, 64, 32\}. 
\vspace{-0.2cm}
\subsection{Results and Discussion}
\subsubsection{\textbf{Overall performance}.} 
We show experiment results in Table \ref{tb:res}, where we find that with the assistance of USR, all methods achieve significant improvements in the CTR and CTCVR tasks within the S\&R scenarios. It demonstrates (1) the effectiveness of fine-grained representations for user interests and scenario-agnostic features in S\&R views; (2) our proposed method GLMT which exploits global labels in the multi-task layer can be adequately trained using the entire set of S\&R label data; (3) the proposed USR is a universal framework that can enhance the performance of many MSL methods in S\&R. This promising property encourages the application of USR to various methods within S\&R.
\vspace{-0.2cm}
\subsubsection{\textbf{Ablation study}.}
\begin{figure}[t!]
\vspace{-0.2cm}
    \centering
    \begin{minipage}[t]{\linewidth}
        \centering
        \includegraphics[width=\columnwidth]{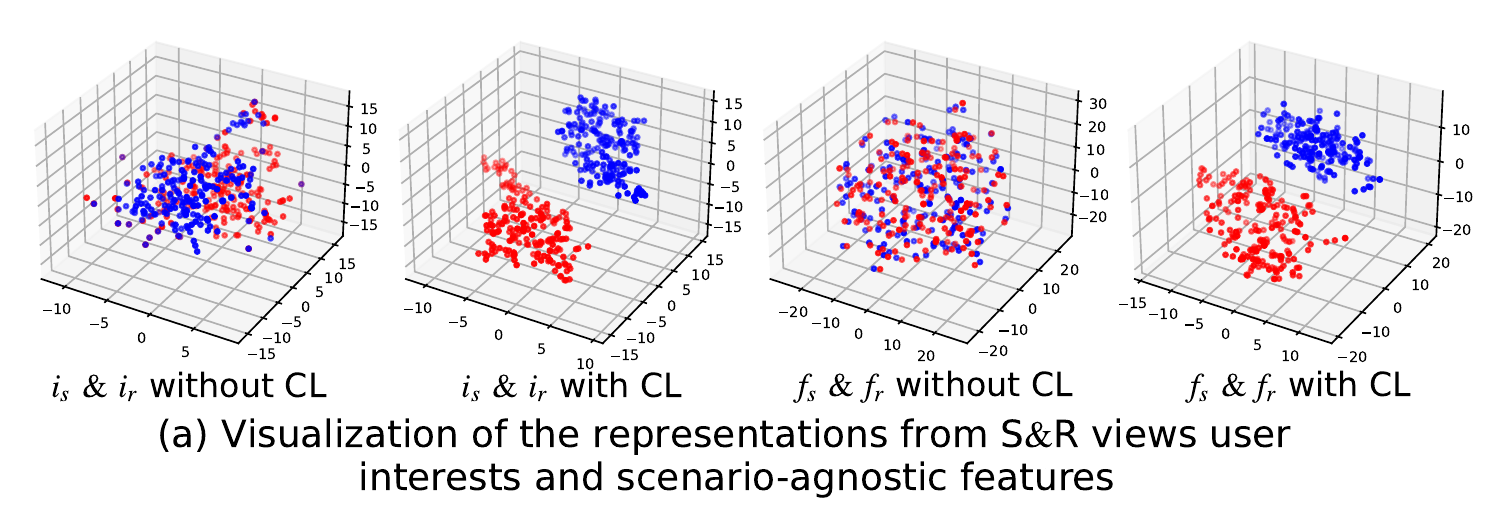}
    \end{minipage}
    \begin{minipage}[t]{\linewidth}
        \centering
        \includegraphics[width=\columnwidth]{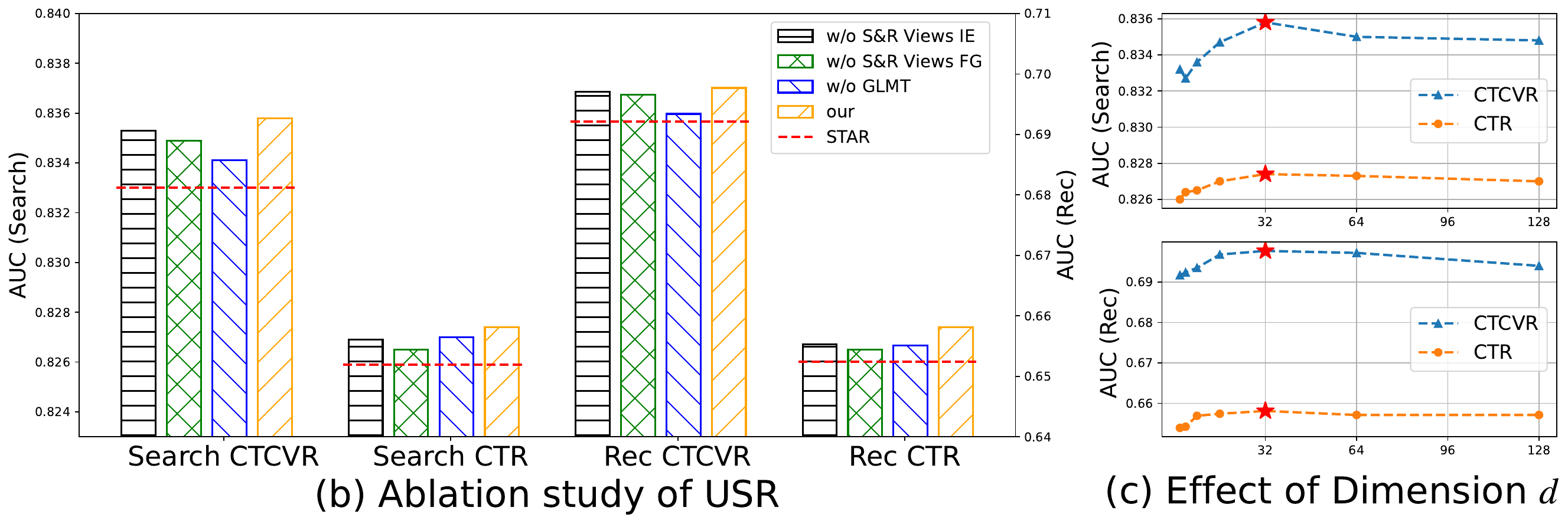}
    \end{minipage}
    \caption{(a) Visualizes the representations from S\&R views user interests and scenario-agnostic features. (b) Shows the ablation study of USR. (c) Shows the effect of  dimension $d$.}
    \label{fig:combined}
\vspace{-0.5cm}
\end{figure}

We perform an ablation study to demonstrate the effectiveness of the different key components. We use STAR \cite{star} as the backbone network for the multi-scenario model. As shown in Fig. \ref{fig:combined} (b),\textbf{ w/o S\&R Views IE} means removing the S\&R Views User Interest Extractor Layer. We can observe that it leads to a significant performance drop, which verifies the importance of extracting and differentiating user interests for S\&R views from global user interests. And \textbf{w/o S\&R Views FG} means that we remove the S\&R Views Feature Generator Layer, and the performance also significantly decreases a lot, indicating the effectiveness of the representations of domain-agnostic features for S\&R views. Finally, w/o \textbf{GLMT} means removing the Global Label Space Multi-Task Layer, which brings the most drop in the performance. This shows that using the entire set of S\&R label data can significantly benefit the joint modeling of S\&R.
\vspace{-0.2cm}
\subsubsection{\textbf{In-depth analyses}.} 
To better understand the benefits of the contrastive loss, we visualize the learned representations of user interests and scenario-agnostic features from S\&R views. Specifically, we use T-SNE \cite{t-sne} to reduce the dimension of the representations to the three-dimensional space and then plot them. The results are shown in Fig. \ref{fig:combined} (a), different colors represent different scenarios. These results indicate that CL can cause representations from the same scenario to cluster closely together, while pushing those from dissimilar scenarios further apart. Additionally, we analyzed the performance impact of varying embedding dimensions $d$ in Fig. \ref{fig:combined} (c), and the results show that $d=32$ yields the best effects for both CTCVR and CTR tasks in S\&R scenarios.
\vspace{-0.2cm}
\subsection{Online A/B Test}
We conducted a one-week online A/B test on the 7Fresh online fresh retail platform. USR contributed to 2.85\% in increase user conversion rate (UCVR) and 1.05\% increase in user click through rate (UCTR) in the search scenario, 3.35\% in increase UCVR and 1.17\% increase in UCTR in the recommendation scenario. The significant improvements achieved by USR confirm its industrial effectiveness. As a result, the proposed USR ranking model has been deployed in the 7Fresh App.
\vspace{-0.2cm}
\section{CONCLUSION}
In this work, we propose an effective and universal S\&R approach to improve Multi-Scenario Learning within S\&R scenarios. We designed two layers to obtain representations of user interests and scenario-agnostic features for S\&R views, which precisely capture the differences within S\&R scenarios. Additionally, we introduce a global label model method in the multi-task layer to fully utilize the entire set of S\&R label data. Experimental results show that with the help of USR, all of the existing MSL models significantly improve, which further promotes the broad applicability of USR. Currently, USR has been fully deployed in the 7Fresh App.
\clearpage

\bibliographystyle{ACM-Reference-Format}

\balance

\bibliography{references}

\end{document}